\documentclass{PoS}

\title{2+1 flavor QCD calculation of $<\mbox{x}>$ and $<\mbox{x}^2>$}

\ShortTitle{2+1 flavor QCD calculation of $<\mbox{x}>$, and $<\mbox{x}^2>$}

\author{$\chi$QCD collaboration: ~\speaker{Devdatta Mankame}, Takumi Doi, Terrence Draper, Keh-Fei Liu  \\
       177 Chem.-Phys. Building, University of Kentucky, 600 Rose Street, Lexington, Kentucky 40506-0055, USA \\  
      E-mail: \email{dmmank2@uky.edu, doi@pa.uky.edu, draper@pa.uky.edu, liu@pa.uky.edu  }}

\author{Thomas Streuer\\
       Institute for Theoretical Physics, University of
Regensburg, 93040 Regensburg, Germany\\
      E-mail: \email{thomas.streuer@physik.uni-regensburg.de}}

\abstract{We calculate the connected insertions of the nucleon three-point function to study the first few moments of the unpolarized structure functions of the nucleon. (The disconnected insertions are discussed elsewhere in these proceedings). The calculation employs the CP-PACS/JLQCD 2+1 dynamical clover fermions on a $16^3\times32$ lattice with lattice spacing $a$=0.1219 fm. The sequential source technique, using non-zero and zero momentum point nucleon field as the secondary source, is applied enabling a study of different currents at various momentum transfer. }

\FullConference{The XXVI International Symposium on Lattice Field Theory \\
                 July 14 - 19, 2008\\
                 Williamsburg, Virginia, USA}
\usepackage{amsmath}
\begin{document}
\section{Introduction}
The calculation of moments of the light cone parton distributions gives detailed insight about the structure of the nucleon. The spin average (unpolarized) structure functions $F_1$ and $F_2$ give information about the density of quarks and gluons in the nucleon, which leads to an understanding of its structure. Structure functions of hadrons can be calculated only non-perturbatively. Using the operator product expansion, it is possible to calculate moments of the structure function. Moments of the quark density distribution are related to the following matrix elements of twist-2 operators \cite{e1,e2,e3,e4,e6} :\nonumber\\
 \begin{eqnarray}
\langle x^{n-1}  \rangle &=& \frac{1}{4} \langle PS|(\frac{i}{2})^{n-1} \bar{\psi} \gamma_{\{\mu_1} \overleftrightarrow{D}_{\mu_2}.... \overleftrightarrow{D}_{\mu_n\}} \psi     |PS\rangle  \label{deri}
\end{eqnarray}

Here \{...\} indicates symmetrization and $\overleftrightarrow{D} = \frac{1}{2} (\overrightarrow{D} - \overleftarrow{D})$. The most promising method known to calculate and understand the quark and gluon structure (i.e. to calculate these matrix elements) is lattice QCD. Lots of calculations have been done in this area in the quenched approximation \cite{e1,e2,e5,e12,e13,e14,e15}, where the vacuum polarization effects are neglected. It is important to calculate various physical quantities for full QCD \cite{e3,e4,e7,e11} (including vacuum polarization effects) and compare these with the quenched ones. 

   We present a calculation for full QCD with relatively heavy pion mass ($\sim$600 MeV). In this calculation we will neglect gluonic operators. We are planning to repeat the same analysis for ``almost physical'' pion mass ($\sim$190 MeV). 
\section{The Lattice approach}

Matrix elements given in Eq. (\ref{deri}) are calculated by evaluating the connected and disconnected insertions as shown in Fig.1. 
To calculate the nucleon matrix elements we first compute two- and three-point correlation functions defined by\\
 \begin{eqnarray}
G_{\Gamma}(t_2,\vec{p}) &=& \sum_{\alpha,\beta} \Gamma_{\alpha,\beta} \langle N_{\beta} (t_2,\vec{p}) \bar{N}_{\alpha} (t_0,\vec{p}) \rangle,\\
G_{\Gamma} (t_2,t_1,\vec{p},{\cal O}) &=& \sum_{\alpha,\beta} \Gamma_{\alpha,\beta} \langle N_{\beta} (t_2,\vec{p}) {\cal O}_{\{\mu_{1}...\mu_{n+1}\}} (t_1) \bar{N}_{\alpha} (t_0,\vec{p}) \rangle 
\end{eqnarray}
where $N$ is the proton interpolating field, given by

\begin{eqnarray}
N_{\alpha} (t_2,\vec{p}) &=& \sum_{\vec{x},a,b,c} e^{-i\vec{p}\cdot\vec{x}} \epsilon_{abc} \psi_{\alpha}^{(u)^a} (x) (\psi^{(u)^b} (x) C\gamma_5 \psi^{(d)^c} (x)) 
\end{eqnarray}
 $C$ is the charge conjugation matrix, $\Gamma$ is the polarization matrix, and
 \begin{eqnarray}
{\mathcal O}_{\{\mu_{1}...\mu_{n+1}\}} &=& \bar{\psi} \gamma_{\{\mu_1} \overleftrightarrow{D}_{\mu_2}...\overleftrightarrow{D}_{\mu_{n+1}\}} \psi     .
\end{eqnarray}

\begin{figure}
\begin{center}
\rotatebox{360}{\includegraphics [width=.4\textwidth, , height=.4\textwidth] {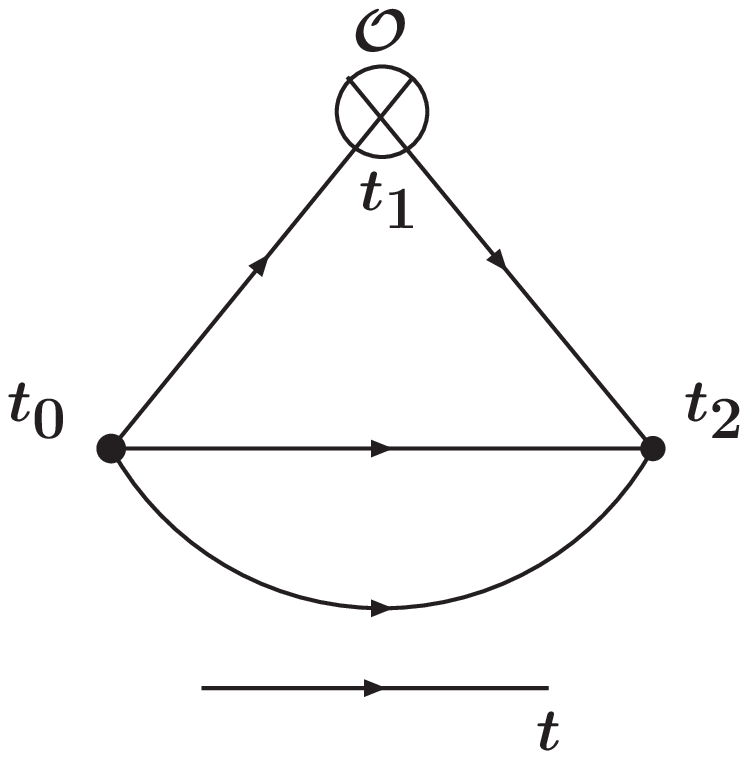}}
\rotatebox{360}{\includegraphics [width=.4\textwidth, , height=.4\textwidth] {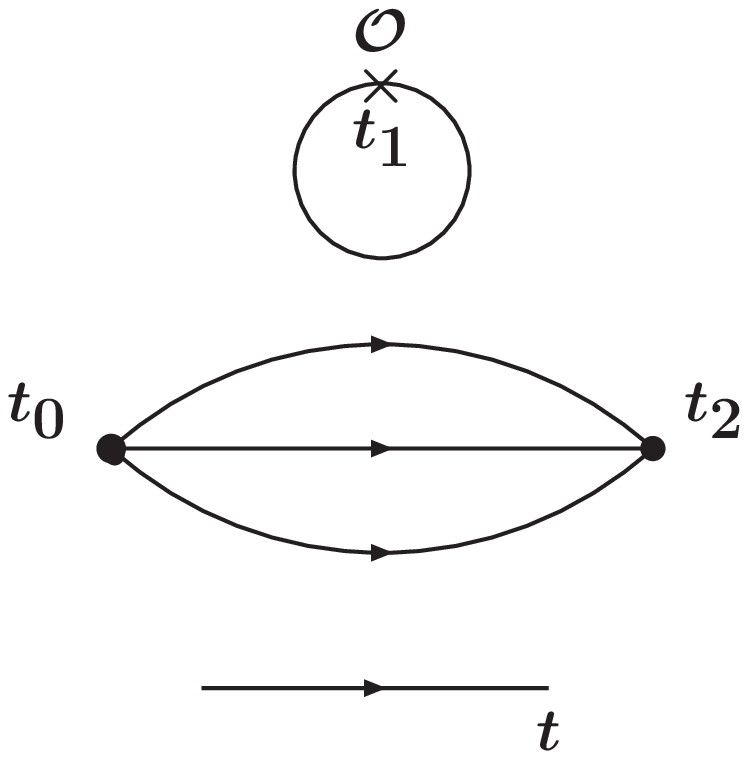}}
\caption{Connected and Disconnected insertions}
\label{fig1}
\end{center}
\end{figure}

To calculate moments of the structure function one needs to take ratios (R) of three-point to two-point functions at zero momentum transfer. These ratios are proportional to the desired matrix elements.

In this work, connected insertions are calculated using the sequential source technique \cite{s16} with fixed sink momentum. This technique fixes the nucleon and its momentum, but leaves the spatial momentum transfer free.

There are some systematic errors we need to take into account. The most prominent systematic errors here are operator mixing, contribution from the excited states, finite size effects, discretization errors, and contributions from the disconnected insertions. (The disconnected insertions are calculated elsewhere \cite{e8}.)

\section{Simulation parameters}

We perform our full QCD calculation for clover fermions with $c_{SW}$=1.7610 and $\kappa_S$=0.13760 on a $16^3\times32$ lattice at lattice spacing {\it{a}}=0.1219 fm. Configurations were taken from the CP-PACS/JLQCD collaboration \cite{e9,e10}. For chiral extrapolation, we run at three different hopping parameters, $\kappa$ = 0.13760 ($m_\pi\sim$800 MeV), 0.13800 ($m_\pi\sim$700 MeV),  and 0.13825 ($m_\pi\sim$600 MeV). The calculation of nucleon matrix elements, in particular derivative operators, requires high statistics. For our calculation, we used 800, 810, and 813 configurations respectively with positive and negative sink momenta ($\pm$1,0,0). We chose polarization along the $z$-direction (for the calculation of magnetic moment and angular momentum).

\section{Results}

Our results for the unpolarized structure functions are listed in Table 1. These are bare (not renormalized) quantities. Results are tabulated for a single source position. We are further planning to take  three different source positions (0,0,0,0), (8,8,8,8) and (16,8,8,16), expecting that this decreases statistical errors by a factor of $\sqrt 3$. Calculations for the remaining two masses are in the pipeline. 

\begin{table}
{\bf{\label1{Operators:}}}
\begin{center}

\begin{tabular}{|c||c||c||r|}
	\hline
\textbf{\em moments}  &  Components   & sink momenta &$\kappa$=0.13760    \\
&&$\vec{p'}$&\\

                  	\hline \hline
$<x>_{u}^{44}$   & ~ $\Theta_{\{44\}}-\frac{1}{3}(\Theta_{\{11\}} + \Theta_{\{22\}} +\Theta_{\{33\}})$ & (0,0,0)&0.397(16) \\
$<x>_{u}^{44}$   & ~ $\Theta_{\{44\}}-\frac{1}{3}(\Theta_{\{11\}} + \Theta_{\{22\}} +\Theta_{\{33\}})$ & ($\pm$1,0,0)&0.436(32) \\
$<x>_{u}^{41}$   & ~ $\Theta_{\{41\}}$  & ($\pm$1,0,0)&0.395(59) \\
\hline
$<x^2>_{u}^{411}$  & ~  $\Theta_{\{411\}}-\frac{1}{2}(\Theta_{\{422\}} +\Theta_{\{433\}})$  & ($\pm$1,0,0) &0.082(31) \\
\hline
$<x>_{d}^{44}$   & ~ $\Theta_{\{44\}}-\frac{1}{3}(\Theta_{\{11\}} + \Theta_{\{22\}} +\Theta_{\{33\}})$ & (0,0,0)  &0.1799(8) \\
$<x>_{d}^{44}$   & ~ $\Theta_{\{44\}}-\frac{1}{3}(\Theta_{\{11\}} + \Theta_{\{22\}} +\Theta_{\{33\}})$ & ($\pm$1,0,0)  &0.191(14) \\
$<x>_{d}^{41}$   & ~ $\Theta_{\{41\}}$ & ($\pm$1,0,0) &0.165(28)\\
\hline
$<x^2>_{d}^{411}$  & ~ $\Theta_{\{411\}}-\frac{1}{2}(\Theta_{\{422\}} +\Theta_{\{433\}})$ & ($\pm$1,0,0) &0.036(16) \\
	\hline
\end{tabular}
\caption{Table 1}
\end{center}
\end{table}

\begin{figure}
\begin{center}
\rotatebox{270}{\includegraphics [width=.4\textwidth, , height=.7\textwidth] {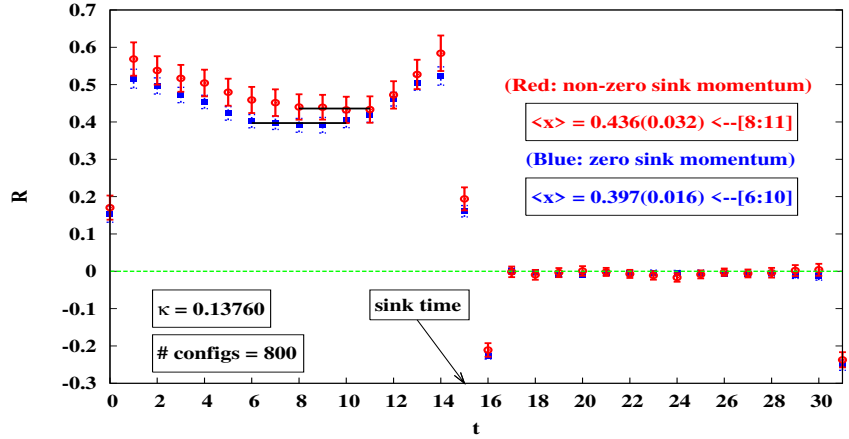}}
\caption{First moment for up quark: $(O_{44}-\frac{1}{3}(O_{11}+O_{22}+O_{33}))$}
\label{fig1}
\end{center}
\end{figure}

\begin{figure}
\begin{center}
\rotatebox{270}{\includegraphics [width=.4\textwidth, , height=.7\textwidth] {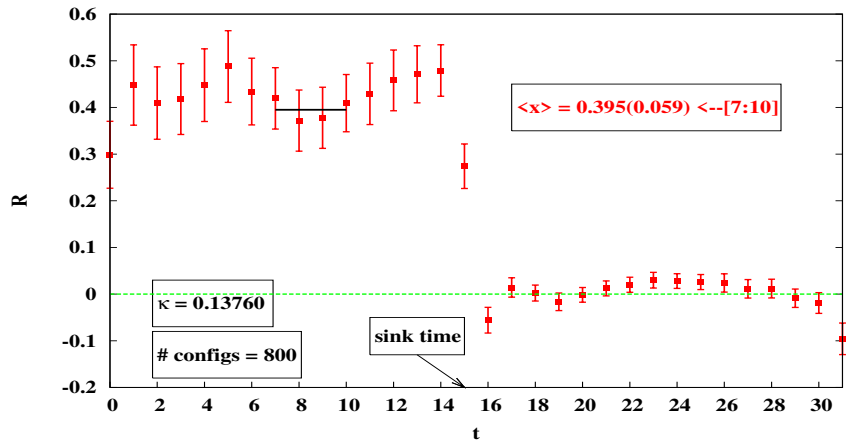}}
\caption{First moment for up quark: $(O_{41})$}
\label{fig1}
\end{center}
\end{figure}

\begin{figure}
\begin{center}
\rotatebox{270}{\includegraphics [width=.4\textwidth, , height=.7\textwidth] {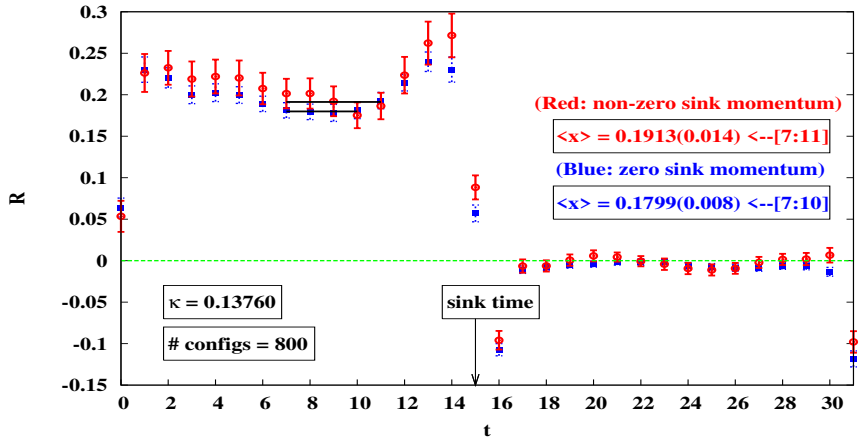}}
\caption{First moment for down quark: $(O_{44}-\frac{1}{3}(O_{11}+O_{22}+O_{33}))$}
\label{fig1}
\end{center}
\end{figure}

\begin{figure}
\begin{center}
\rotatebox{270}{\includegraphics [width=.4\textwidth, , height=.7\textwidth] {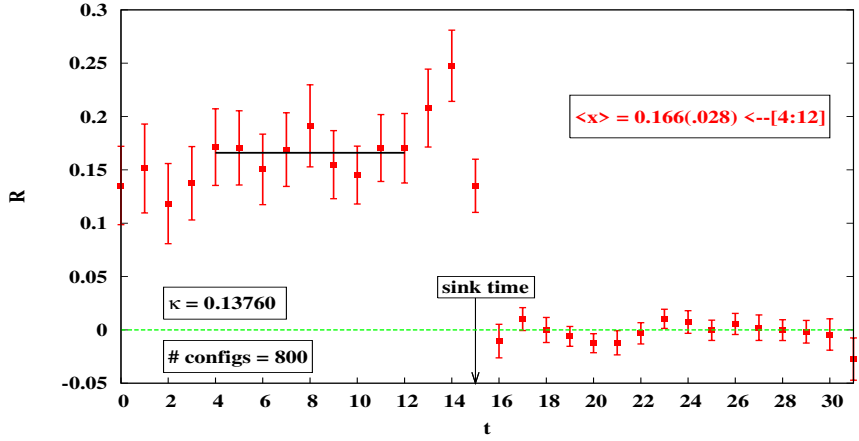}}
\caption{First moment for down quark: $(O_{41})$}
\label{fig1}
\end{center}
\end{figure}

\begin{figure}
\begin{center}
\rotatebox{270}{\includegraphics [width=.4\textwidth, , height=.7\textwidth] {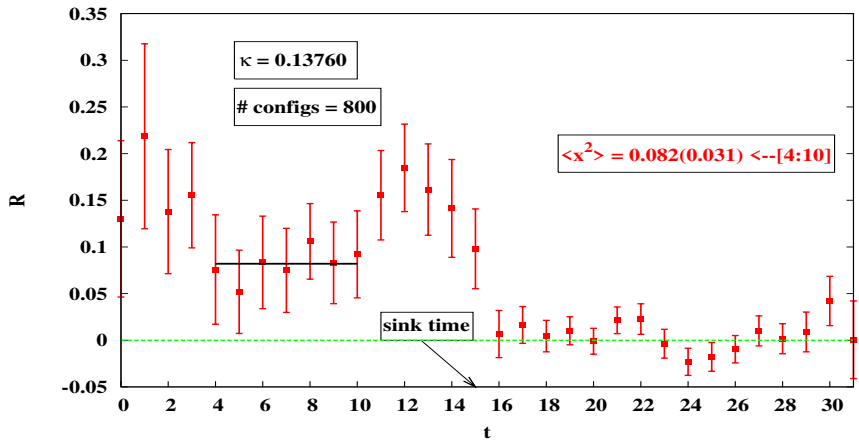}}
\caption{Second moment for up quark: $(O_{411}-\frac{1}{2}(O_{422} + O_{433}))$}
\label{fig1}
\end{center}
\end{figure}

\begin{figure}
\begin{center}
\rotatebox{270}{\includegraphics [width=.4\textwidth, , height=.7\textwidth] {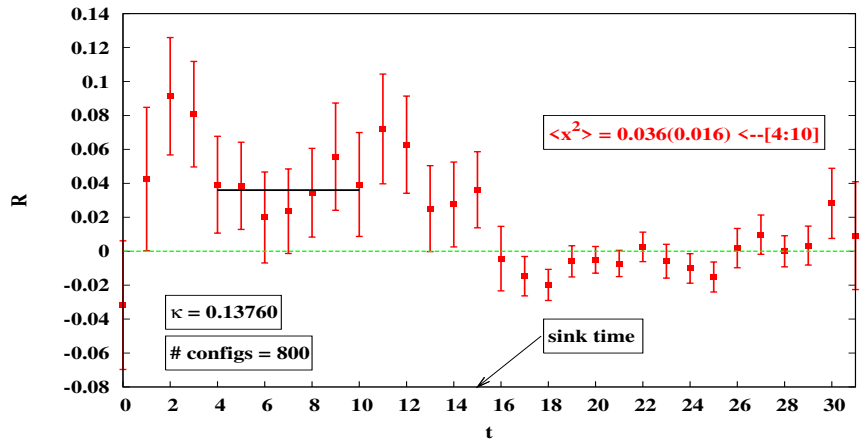}}
\caption{Second moment for down quark: $(O_{411}-\frac{1}{2}(O_{422} + O_{433}))$}
\label{fig1}
\end{center}
\end{figure}

\section{Conclusion}
We have calculated $\langle x \rangle$ and $\langle x^2 \rangle$ for the unpolarized structure function. The results so far are encouraging. In order to compare our results to the experimental ones, we have to renormalize them, taking into account operator mixing.

For the larger lattice, we are planning to include smearing at both sink and source to reduce contamination from the excited states. We may incorporate operator improvement to reduce the discretization effect. The large lattice size will reduce finite size effects significantly.

 We are also working on electric and magnetic form factors for various momentum transfer and quark angular momentum. 

\section{Acknowledgments}
We would like to thank Dr. B. Doyle and Dr. M. Deka for technical help. Numerical calculations for the present work have been done at the JLab supercomputing facility and partially at Franklin (NERSC) and BCX (University of Kentucky).

\end{document}